\documentclass[letterpaper, 10 pt, conference]{ieeeconf}

\IEEEoverridecommandlockouts                              

\usepackage{cite}
\usepackage{grffile}
\usepackage{caption}
\usepackage{graphicx}
\usepackage{amssymb,amsmath}
\usepackage{epsfig}
\usepackage{epstopdf}
\usepackage{setspace}
\usepackage{amsthm}
\usepackage{enumerate}
\usepackage{tikz}
\usepackage{multirow}
\usetikzlibrary{shapes,arrows}
\title{\LARGE \bf
Distributed Nash Equilibrium Seeking By Gossip in Games on Graphs
}

\author{Farzad Salehisadaghiani, and Lacra Pavel
\thanks{The authors are with the Department of Electrical and Computer Engineering, University of Toronto, Toronto, ON M5S 3G4, Canada (e-mails: {\tt\small farzad.salehisadaghiani@mail.utoronto.ca, pavel@ece.utoronto.ca}).}}
\newtheorem{assumption}{Assumption}
\newtheorem{lemma}{Lemma}
\newtheorem{theorem}{Theorem}
\newtheorem{remark}{Remark}

\newtheorem{definition}{Definition}

\newtheorem{corollary}{Corollary}

\begin{document}
\allowdisplaybreaks

\maketitle
\thispagestyle{empty}
\pagestyle{empty}

\begin{abstract}
We consider a gossip approach for finding a Nash equilibrium in a distributed multi-player network game. We extend previous results on Nash equilibrium seeking to the case when the players' cost functions may be affected by the actions of any subset of players. An interference graph is employed to illustrate the partially-coupled cost functions and the asymmetric information requirements. For a given interference graph, we design a generalized communication graph so that players with possibly partially-coupled cost functions exchange only their required information and make decisions based on them. Using a set of standard assumptions on the cost functions, interference and communication graphs, we prove almost sure convergence to a Nash equilibrium for diminishing step sizes. We then quantify the effect of the second largest eigenvalue of the expected communication matrix on the convergence rate, and illustrate the trade-off between the parameters associated with the communication and the interference graphs. Finally, the efficacy of the proposed algorithm on a large-scale networked game is demonstrated via simulation.
\end{abstract}

\section{INTRODUCTION}
Distributed seeking of Nash equilibria in networked games has received considerable attention in recent years \cite{stankovic2012distributed, yin2011nash, frihauf2012nash, jayash10, li2010competitive, chen2012spatial, alpcan2004hybrid, Marden2013}. A networked game can be represented by a \emph{graphical model} which enables us to index the cost function of each player as a function of player's own actions and those of his neighbors in the graph. There are many real-world applications that motivate us to generalize the Nash seeking problem to a graphical game setup \cite{li2010competitive}, \cite{chen2012spatial}. The collection of transmitters and receivers in a wireless data network can be captured as a graphical model. Interferences among the transmitters and receivers affect the players' signal-to-interference ratio (SIR) \cite{alpcan2004hybrid}. An optical network is another relevant application that can be modeled as a graphical game. The channels are assumed to be the players and interferences, which affect the optical signal-to-noise ratio (OSNR) of each channel, can be modeled by graph edges, \cite{Pavelbook2012}.

In this work we design a locally distributed algorithm for Nash equilibrium seeking in a graphical game. In such a game, the players' cost functions may be affected by the actions of \emph{any subset} of players. They exchange the required information locally according to a communication graph and update their actions to optimize their cost functions. Due to limited information available from local neighbors, each player maintains an estimate of the other players' actions and update their estimates during the iterations.

\emph{Literature review.} A graphical game is a succinct representation of a multi-player game considering the local interactions and the sparsity of the interferences. A graphical game can be simply described by an undirected graph called \emph{interference graph} in which the players are marked by the vertices and the interferences are represented by the edges \cite{nisan2007algorithmic}, \cite{kearns2001graphical}.

The idea of a graphical game has been used in various areas. In congestion games, \cite{tekin2012atomic} considers a generalization to graphical games. The model involves the spatial positioning of the players which affects their performances. A \emph{conflict graph} is defined to specify the players that cause congestion to each other. A methodology is presented in \cite{Marden2013} for games with local cost functions which are dependent on information from only a set of local neighboring agents. Extra state space variables are defined for the game to achieve a desired degree of locality. For dynamical games, graphical games are considered where the dynamic of each player depends only on local neighbor information \cite{abouheaf2014multi}. A stronger definition of interactive Nash equilibrium is used to guarantee a unique Nash equilibrium. The information flow is restricted by a communication graph which is identical to the interference graph. In an economic setting, \cite{bramoulle2014strategic} draws attention to the problem of "who interacts with whom" in a network. This paper states the importance of communication with neighboring players in the network. The effect of local peers on increasing the usage level of consumers is addressed in \cite{candogan2012optimal}. Using \emph{word-of-mouth} communication, the players typically form their opinions about the quality of a product and improve their purchasing behavior according to the information they obtain from their local peers.

For generalized convex games the problem of finding a Nash equilibrium is studied in \cite{zhu2016distributed}. The communication graph is \emph{identical} to the interference graph. A connected communication graph is considered in \cite{Jayash} for aggregative games. For a large class of convex games, \cite{Salehisadaghiani2014Nash} proposes an asynchronous gossip-based algorithm over a connected communication graph for a \emph{complete interference graph}.

In this work, we generalize the algorithm in \cite{Salehisadaghiani2014Nash} to the case where the interference graph is not a complete graph, i.e., the players' cost functions may depend on the actions of any "subset" of players. A communication graph, which is a subset of the interference graph, is designed for the network. We prove that there is lower bound for the communication graph under which the algorithm converges to a Nash equilibrium for diminishing step sizes. We then investigate the convergence rate of the algorithm. The results show a trade-off between the parameters associated with the communication graph and the ones associated with the interference graph. 

Due to space limitations most of the proofs are omitted.
\subsection{Graph Theory Notions}
The following definitions are from \cite{godsil2013algebraic}, \cite{goddard1993note}. A subgraph $H$ of a graph $G$ is a graph whose vertices and edges are a subset of the vertex set and edge set of $G$, respectively. A subgraph $H$ is a spanning subgraph of $G$, if it contains all the vertices of $G$. A triangle-free subgraph $H$ of a graph $G$ is a subgraph in which no three vertices form a triangle of edges. $H$ is a maximal triangle-free subgraph of $G$ if adding a missing edge to $H$ creates a triangle.
\section{Problem Statement}\label{problem_statement}

Consider a multi-player game in a network with a set of players $V=\{1,\ldots,N\}$. For $i\in V$, there is a real-valued function $J_i$ indicating player $i$'s individual cost function. The players' cost functions are not necessarily fully coupled in the sense that they may be affected by the actions of any number of other players. To illustrate the partially coupled cost functions, we define an \emph{interference graph}, denoted by $G_I(V,E_I)$, with $E_I$ marking player pairs that interfere one with another. We denote with $N_I(i)$, the set of neighbors of player $i$ in $G_I$, i.e., $N_I(i):=\{j\in V|(i,j)\in E_I\}$. We also define $\tilde{N}_I(i):=N_I(i)\cup\{i\}$.
\begin{assumption}\label{connected_undirected}
$G_I$ is connected and undirected.
\end{assumption}
Let $\Omega_j\subset\mathbb{R}$ denote the action set of player $j$. We denote by $\Omega$ the action set of all players, i.e., $\Omega=\prod_{i\in V}\Omega_i\subset\mathbb{R}^N$ where $\prod$ denotes the Cartesian product. For $i\in V$, $J_i:\Omega^i\rightarrow \mathbb{R}$ is the cost function of player $i$ where $\Omega^i=\prod_{j\in\tilde{N}_I(i)}\Omega_j\subset\mathbb{R}^{|\tilde{N}_I(i)|}$ is the action set of players affecting the cost function of player $i$. The game denoted by $\mathcal{G}(V,\Omega_i,J_i, G_I)$ is defined based on the set of players $V$, the action set $\Omega_i$, $\forall i\in V$, the cost function $J_i$, $\forall i\in V$ and $G_I$.

For $i\in V$, let $x^i=(x_i,x_{-i}^i)\in\Omega^i$, with $x_i\in\Omega_i$ and $x_{-i}^i\in\Omega_{-i}^i:=\prod_{j\in N_I(i)}\Omega_j$, denote the other players' actions which affect the cost function of player $i$. Let also $x=(x_i,x_{-i})\in\Omega$, with $x_i\in\Omega_i$ and $x_{-i}\in\Omega_{-i}:=\prod_{j\in V/\{i\}}\Omega_j$, denote all other players' actions except $i$.

The game defined on $G_I$ is played such that for given $x_{-i}^i\in \Omega_{-i}^i$, each player $i$ aims to minimize his own cost function selfishly to find an optimal action,
\begin{equation}
\label{mini_0}
\begin{aligned}
& \underset{y_i}{\text{minimize}}
& & J_i(y_i,x_{-i}^i) \\
& \text{subject to}
& & y_i\in \Omega_i.
\end{aligned}
\end{equation}

Note that there are $N$ separate simultaneous optimization problems and each of them is run by a particular player $i$. We assume that the cost function $J_i$ and the action set $\Omega^i$ are only available to player $i$.  Thus every player knows which other players' actions affect his cost function.

A Nash equilibrium of the game for the case when $G_I$ is not a complete graph is defined as follows.
\begin{definition}\label{Nash_def}
Consider an $N$-player game $\mathcal{G}(V,\Omega_i,J_i, G_I)$, each player $i$ minimizing the cost function $J_i:\Omega^i\rightarrow\mathbb{R}$. A vector $x^*=(x_i^*,x_{-i}^*)\in\Omega$ is called a Nash equilibrium of this game if for every given ${x_{-i}^i}^*\in\Omega_{-i}^i$
\begin{equation}
J_i(x_i^*,{x_{-i}^{i\,*}})\leq J_i(x_{i},{x_{-i}^{i\,*}})\quad\forall x_i\in \Omega_i,\,\,\forall i\in V.
\end{equation}
\end{definition}
The players are required to exchange some information to update their actions. A \emph{communication graph} $G_C(V,E_C)$ is defined with $E_C\subseteq V\times V$ denoting the set of communication links between the players. $(i,j)\in E_C$ if and only if players $i$ and $j$ communicate. The set of neighbors of player $i$ in $G_C$, denoted by $N_C(i)$, is defined as $N_C(i):=\{j\in V|(i,j)\in E_C\}$. In order to reduce the number of communications between players, we design a $G_C$ such that only the required information is exchanged.
\begin{assumption}\label{connectivity}
The communication graph $G_C$ satisfies
\begin{itemize}
  \item $G_m\subseteq G_C\subseteq G_I$, if $G_I$ has a maximal triangle-free spanning subgraph, $G_m$,
  \item $G_C=G_I$, otherwise.
\end{itemize}
\end{assumption}

A Nash equilibrium can be efficiently computed by solving a variational inequality $VI(\Omega, F)$ where $\Omega\subset\mathbb{R}^N$ is the action set of all players and $F:\Omega\rightarrow\mathbb{R}^N$ is a pseudo-gradient mapping defined as $F(x):=[\nabla_{x_i}J_i(x^i)]_{i\in V}$ \cite{Facchi24}.

In the following we state a few assumptions for the existence and the uniqueness of a Nash equilibrium.
\begin{assumption}
\label{assump}
For every $i\in V$, the action set $\Omega_i$ is a non-empty, compact and convex subset of $\mathbb{R}$. $J_i(x_i,x_{-i}^i)$ is a continuously differentiable function in $x_i$, jointly continuous in $x^i$ and convex in $x_i$ for every $x_{-i}^i$.
\end{assumption}

The compactness of $\Omega$ implies that $\forall i\in V$ and $x^i\in\Omega^i$, 
\begin{equation}\label{bounded}
\|\nabla_{x_i}J_i(x^i)\|\leq C,\quad\text{for some }C>0.
\end{equation}
\begin{assumption}\label{Lip_assump}
$F:\Omega\rightarrow\mathbb{R}^N$ is strictly monotone,
\begin{equation}
(F(x)-F(y))^T(x-y)> 0\quad\forall x,y\in \Omega,\text{ }x\neq y.
\end{equation}
\end{assumption}
\begin{assumption}\label{Lip_assump2}
$\nabla_{x_i}J_i(x_i,u)$ is Lipschitz continuous in $x_i$, for every fixed $u\in\Omega_{-i}^i$ and for every $i\in V$, i.e., there exists $\sigma_i>0$ such that
\begin{equation}
\|\!\nabla_{x_i}\!J_i\!(x_i,\!u)\!-\!\nabla_{x_i}\!J_i\!(y_i,\!u)\!\|\!\leq\! \sigma_i\|x_i-y_i\|\quad\forall x_i,y_i\in\Omega_{i}.
\end{equation}

Moreover, $\nabla_{x_i}J_i(x_i,u)$ is Lipschitz continuous in $u$ with a Lipschitz constant $L_i>0$ for every fixed $x_i\in\Omega_i,\,\forall i\in V$.
\end{assumption}

Our objective is to find an algorithm for computing a Nash equilibrium of $\mathcal{G}(V,\Omega_i,J_i, G_I)$ with partially coupled cost functions as described by $G_I(V,E_I)$ using only imperfect information over the communication graph $G_C(V,E_C)$.
\section{Asynchronous Gossip-based Algorithm}\label{asynch}
We propose a distributed algorithm, using an asynchronous gossip-based method in \cite{Salehisadaghiani2014Nash}. We obtain a Nash equilibrium of $\mathcal{G}(V,\Omega_i,J_i, G_I)$ by solving the associated $VI$ problem by a projected gradient-based approach with diminishing step size.
The mechanism of the algorithm can be briefly explained as follows:
Each player builds and maintains an estimate $\hat{x}_j^i$, $j\in \tilde{N}_I(i)$ of the actions which affect his cost function (as in $G_I$) and locally communicates with his neighbors over $G_C$ to exchange his estimates and update his action.

The algorithm is elaborated in the following steps:\\
1- \textbf{\emph{Initialization Step:}}
Each player $i$ maintains an initial \emph{temporary} estimate for the players whose actions affect his cost function, $\tilde{x}_j^i(0)\in\Omega_j\subset\mathbb{R}$, $j\in\tilde{N}_I(i)$.\\
2- \textbf{\emph{Gossiping Step:}}
At the gossiping step, player $i_k$ wakes up at $T(k)$ and selects a communication neighbor indexed by $j_k\in N_C(i_k)$.
They exchange their temporary estimate vectors and construct their estimate of the players whose actions affect their cost functions.

The estimates are computed as in the following:
\begin{eqnarray}\label{excluding}
\hspace{-0.15cm}\text{1)} \begin{cases}
\hat{x}_{l}^{i_k}(k)=\frac{\tilde{x}_{l}^{i_k}(k)+\tilde{x}_{l}^{j_k}(k)}{2},& l\in (N_I(i_k)\cap\tilde{N}_I(j_k)) \\ \hat{x}_{l}^{j_k}(k)=\frac{\tilde{x}_{l}^{i_k}(k)+\tilde{x}_{l}^{j_k}(k)}{2},& l\in (N_I(j_k)\cap\tilde{N}_I(i_k)).
\end{cases}
\end{eqnarray}
\begin{eqnarray}\label{including}
\hspace{-0.2cm}\text{2)} \begin{cases}
\hat{x}_{r}^{i_k}\!(\!k)\!=\!\tilde{x}_{r}^{i_k}\!(\!k),&r\in \tilde{N}_I(i_k)\backslash (N_I(i_k)\cap\tilde{N}_I(j_k))\\
\hat{x}_{r}^{j_k}\!(\!k)\!=\!\tilde{x}_{r}^{j_k}\!(\!k),&r\in \tilde{N}_I(j_k)\backslash (N_I(j_k)\cap\tilde{N}_I(i_k)).
\end{cases}
\end{eqnarray}\\
3) For all other $i\notin \{i_k,j_k\}$,
\begin{equation}\label{other_inc_exc}
\hat{x}_j^i(k)=\tilde{x}_j^i(k),\quad\forall i\notin \{i_k,j_k\},\,\forall j\in\tilde{N}_I(i).
\end{equation}
Note that the player $i$'s estimate of his action is indeed his action, i.e., $\tilde{x}_i^i(k)=x_i(k)$ for all $i\in V$.\\
3- \textbf{\emph{Local Step:}}
All the players update their actions according to a projected gradient-based method. Let $\hat{x}^i=(\hat{x}_i^i,\hat{x}_{-i}^i)\in\Omega^i$, with $\hat{x}_i^i\in\Omega_i$ as player $i$'s estimate of his action and $\hat{x}_{-i}^i\in\Omega_{-i}^{i}$ as the estimate of the players whose actions affect player $i$'s cost function. Player $i$ updates his action as follows: if $i\in\{i_k,j_k\}$,
\begin{equation}\label{local_step}
x_i(k+1)=T_{\Omega_i}[x_i(k)-\alpha_{k,i} \nabla_{x_i}J_i(x_i(k),\hat{x}_{-i}^i(k))],
\end{equation}
otherwise, $x_i(k+1)=x_i(k)$. In \eqref{local_step}, $T_{\Omega_i}:\mathbb{R}\rightarrow\Omega_i$ is an Euclidean projection and $\alpha_{k,i}$ are diminishing step sizes such that $\sum_{k=1}^{\infty}\alpha_{k,i}^2<\infty$, $\sum_{k=1}^{\infty}\alpha_{k,i}=\infty$ $\forall i\in V$. Note that $\alpha_{k,i}$ is inversely related to the number of updates $\nu_k(i)$ that each player $i$ has made until time $k$ (i.e., $\alpha_{k,i}=\frac{1}{\nu_k(i)}$).

At this moment all the temporary estimates are updated for every $i\in V,\,j\in\tilde{N}_I(i)$ as follows:
\begin{eqnarray}\label{temp_update}
\hspace{-1.05cm}\tilde{x}_j^i(k+1)=\begin{cases} \hat{x}_j^i(k),&\text{if }i\neq j \\ x_i(k+1),& \text{if }i=j.\end{cases}\end{eqnarray}

In \eqref{temp_update} for $j=i$, player $i$'s temporary estimate is updated by his action.
At this point, the players begin a new iteration from step 2. 

The algorithm is inspired by \cite{Salehisadaghiani2014Nash} except that only the required information is exchanged. When $G_I$ is not complete, the proposed algorithm can offer substantial savings. 
\section{Convergence For Diminishing Step Size}\label{convergence_diminish}
Consider a memory $\mathcal{M}_k$ to denote the {\emph{sigma-field}} generated by the history up to time $k-1$ with $\mathcal{M}_0=\mathcal{M}_1=\{\tilde{x}^i(0),\text{ }i\in V\}$.
$$\mathcal{M}_k=\mathcal{M}_0\cup\Big\{(i_l,j_l); 1\leq l\leq k-1\Big\},\quad \forall k\geq 2.$$

For player $i$, let $m_i:=\text{deg}(i)+1$ where deg$(i)$ is the degree of vertex $i\in V$ in $G_I$. Let also $m:=\sum_{i=1}^{N}m_i$ and $\textbf{m}:=[m_1,\ldots,m_N]^T\in\mathbb{R}^N$.
\begin{remark}\label{mi>1}
Assumption~\ref{connected_undirected} implies that $m_i>1$, $\forall i\in V$ and $m>N$. If $G_I$ is a complete graph (or in other words if the cost functions are fully coupled), then $m=N^2$.
\end{remark}
The following lemma holds for $G_I$ and $G_C$.
\begin{lemma}\label{tria_lemma}
Let $G_I$ and $G_C$ satisfying Assumptions~\ref{connected_undirected}, \ref{connectivity}. Then every estimate is exchanged after sufficiently many iterations.
\end{lemma}

In the following we write the algorithm in a compact form. Let $B=A+I_N\in \mathbb{R}^{N\times N}$, where $A=[a_{ij}]_{i,j\in V}$ is the adjacency matrix associated with $G_I$ with $a_{ij}=1$ if $(i,j)\in E_I$ and $a_{ij}=0$ otherwise. Let also
\begin{equation}\label{s_ij}
s_{ij}:=\sum_{l=1}^{j}B(i,l)+\delta_{i\neq 1}\sum_{r=1}^{i-1}m_r,
\end{equation}
where $\delta_{i\neq 1}=1$ if $i\neq 1$ and $\delta_{i\neq 1}=0$ if $i=1$. For each pair $i,j\in V$, we assign a vector $E_j^i\in \mathbb{R}^m$.
\begin{eqnarray}\label{E_j^i}
E_j^i=
\begin{cases}
e_{s_{ij}},&\text{if }i\in V,\,j\in\tilde{N}_I(i)\\
\textbf{0}_{m},&\text{if }i\in V,\,j\notin\tilde{N}_I(i),
\end{cases}
\end{eqnarray}
where $e_i$ is a unit vector in $\mathbb{R}^m$ whose $i$-th element is 1 and $\textbf{0}_m$ is the all zeros vector in $\mathbb{R}^m$.

The communication matrix $W(k)$ is defined as
\begin{equation}\label{W_def}
  W(k):=I_m-\frac{1}{2}\sum_{l\in\text{ind}(i_k,j_k)}(E_l^{i_k}-E_l^{j_k})(E_l^{i_k}-E_l^{j_k})^T,
\end{equation}
where $\text{ind}(i_k,j_k):=\{z\in V : B(i_k,z)\cdot B(j_k,z)=1\}$ is the set of indices that belong to $\tilde{N}_I(i_k)\cap\tilde{N}_I(j_k)$ for $i_k,\,j_k\in N_C$.
\begin{remark}\label{barabari_W_ha}
$W(k)$ is a generalized communication matrix. For the fully coupled case (i.e., $\tilde{N}_I(i)=V$), it reduces to the definition in \cite{Salehisadaghiani2014Nash}, i.e., $W(k)=\big(I_N-(e_{i_k}-e_{j_k})(e_{i_k}-e_{j_k})^T\big)\otimes I_N$ where $e_i$ is a unit vector in $\mathbb{R}^N$.
\end{remark}

The definition \eqref{W_def} implies that $W(k)$ is a doubly stochastic matrix such that $W(k)^T\mathbf{1}_m=W(k)\mathbf{1}_m=\mathbf{1}_m$ where $\mathbf{1}_m$ is the all ones vector in $\mathbb{R}^m$.

Let $\tilde{x}(k):=\big[\tilde{x}^{1^T},\ldots,\tilde{x}^{N^T}\big]^T$ be the stack vector of all temporary estimates and $\bar{x}(k)=W(k)\tilde{x}(k)$. Then, $\hat{x}_{-i}^i(k)=[\bar{x}_{r}(k)]_{r\in I(i)}$, where $I(i):=\{d:d=s_{ij},\ j\in N_I(i)\}$ and $s_{ij}$ is as in \eqref{s_ij}.

The convergence proof has two steps:
\begin{enumerate}
  \item First, we prove almost sure convergence of the temporary estimate vector to the average of all temporary estimate vectors (Theorem~1).
  \item Secondly, we prove convergence of the actions toward the Nash equilibrium, almost surely (Theorem~2).
\end{enumerate}

The average of all temporary estimates of the players is denoted by $z(k)\in\mathbb{R}^N$. Let $z(k):=\bar{H}\tilde{x}(k)\in\mathbb{R}^N$ where,
\begin{eqnarray}
\label{Hbar}\bar{H}&:=&\text{diag}(1./\textbf{m})H^T\in\mathbb{R}^{N\times m},\\
1./\textbf{m}&:=&[\frac{1}{m_1},\ldots,\frac{1}{m_N}]^T,\nonumber\\
\label{H}H&:=&[\sum_{i=1}^{N}E_1^i,\ldots,\sum_{i=1}^{N}E_N^i]\in\mathbb{R}^{m\times N}.
\end{eqnarray}
The augmented average of all temporary estimates is as follows:
\begin{equation}\label{ave_Z}
Z(k):=Hz(k)=H\bar{H}\tilde{x}(k).
\end{equation}

The convergence proof depends on some key properties of $W$ and $H$ given in Lemma~\ref{lemma_ext_stoch}-\ref{gamma_less_1}.
\begin{lemma}\label{lemma_ext_stoch}
Let W(k) and $H$ be defined in \eqref{W_def} and \eqref{H}. The following properties hold:
\begin{eqnarray}\label{ext_stoch}
W^T(k)W(k)=W(k),\,W(k)H=H,\,H^TW(k)=H^T.\nonumber
\end{eqnarray}
\end{lemma}

\begin{lemma}\label{QZ=0}
Let $Q(k)\!:=\!W\!(k)\!-\!H\!\bar{H}\!W\!(k)$. Then $Q(k)\!Z\!(k)\!=\!0$.
\end{lemma}

\begin{lemma}\label{R=1}
Let $R:=I_m-H\bar{H}$ with $H$, $\bar{H}$ defined in \eqref{H}, \eqref{Hbar}. Then $\|R\|=1$, where $\|\cdot\|$ denoting the reduced norm.
\end{lemma}
\begin{lemma}\label{gamma_less_1}
Let $Q(k):=W(k)-H\bar{H}W(k)$ and $\gamma=\lambda_{\max}\big(\mathbb{E}[Q(k)^TQ(k)]\big)$. Then $\gamma<1$.
\end{lemma}

Using Lemmas~\ref{lemma_ext_stoch}-\ref{gamma_less_1}, we show in the following theorem that under Assumptions~\ref{connected_undirected}-\ref{assump} $\tilde{x}(k)$ converges to $Z(k)$.
\begin{theorem}\label{consensus1}
Let $\tilde{x}(k)$ be the stack vector with all temporary estimates and $Z(k)$ be its average as in \eqref{ave_Z}. Let also $\alpha_{k,\text{max}}=\max_{i\in V}\alpha_{k,i}$. Then under Assumptions~\ref{connected_undirected}-\ref{assump},
\begin{enumerate}[i)]
\item $\sum_{k=0}^{\infty}\alpha_{k,\text{max}}\|\tilde{x}(k)-Z(k)\|<\infty$,
\item $\sum_{k=0}^{\infty}\|\tilde{x}(k)-Z(k)\|^2<\infty$.
\end{enumerate}
\end{theorem}

Theorem~\ref{consensus1} yields the following corollary.
\begin{corollary}\label{remark}
Let $z(k):=\bar{H}\tilde{x}(k)\in\mathbb{R}^N$ be the average of all players' temporary estimates. Under Assumptions~\ref{connected_undirected}-\ref{assump}, the following hold for players' actions $x(k)$.
\begin{enumerate}[i)]
\item $\sum_{k=0}^{\infty}\alpha_{k,\text{max}}\|x(k)-z(k)\|<\infty$,
\item $\sum_{k=0}^{\infty}\|x(k)-z(k)\|^2<\infty.$
\end{enumerate}
\end{corollary}

By Theorem~\ref{consensus1} and Corollary~\ref{remark}, $\tilde{x}(k)$ and $x(k)$ converge to $Z(k)$ and $z(k)$, respectively as $k\rightarrow\infty$.
\begin{theorem}
\label{Prop_algo_1}
Let $x(k)$ and $x^*$ be all players' actions and the Nash equilibrium of $\mathcal{G}$, respectively. Under Assumptions~\ref{connected_undirected}-\ref{Lip_assump2}, the sequence $\{x(k)\}$ generated by the algorithm converges to $x^*$, almost surely.
\end{theorem}

\section{convergence rate}\label{convergence_rate}

In this section we compare the convergence rate of the algorithm proposed in Section~\ref{asynch} (denoted as Algorithm~1) with the algorithm in \cite{Salehisadaghiani2014Nash} (denoted as Algorithm~2). Algorithm~1 is an extension of Algorithm~2 by considering partially-coupled cost functions for the players via $G_I$. Algorithm~2 operates as if a fully-coupled cost function is assigned to each player and the interference graph is complete.

By Assumption~\ref{connectivity}, any feasible communication graph for Algorithm~1 has a lower bound $G_m$, however, the communication graph for Algorithm~2 can be minimally connected. Thus, for the best case scenario we expect more iterations for Algorithm~1 than Algorithm~2 from the point of view of the parameters associated with $G_C$.

In the following, we compare Algorithm~1 and Algorithm~2 from the point of view of $G_I$. We can show that for each iteration, Algorithm~1 needs less time than Algorithm~2 since less information (estimate) is needed to be exchanged.

For the sake of comparison, we assume that both algorithms run over the same $G_C\supseteq G_m$. Let $r$ be the time required to exchange an estimate, and let $s$ be the time required to process a full gradient. Note that the processing time for the gradient is linearly dependent on the data set. We ignore the time required to compute the projection in the local step. Thus for each iteration, the average time required to exchange all the estimates between players $i$ and $j$ and to update the actions under Algorithm~1 is
\begin{equation}\label{average_time_interference}
T_\text{av}^1:=\sum_{i\in V}\sum_{j\in N_C(i)}\frac{1}{N}p_{ij}\Big(|N_I(i)\cap N_I(j)|r+\frac{m_i}{N}s\Big),
\end{equation}
where $p_{ij}$ is the probability that players $i$ and $j$ contact each other. $|N_I(i)\cap N_I(j)|r$ is the time required to exchange all the estimates of player $i$ which affect player $j$'s cost function except player $i$'s action. $\frac{m_i}{N}s$ is the time required to compute $\nabla_{x_i}J_i(x_i,x_{-i}^i)$, noting that $s$ is the processing time for computing $\nabla_{x_i}J_i(x_i,x_{-i})$.

In Algorithm~2 the average time for each iteration is computed by replacing $|N_I(i)\cap N_I(j)|$ and $m_i$ in \eqref{average_time_interference} with $N-1$ and $N$, respectively. Then we obtain,
\begin{equation}\label{average_time_communication}
T_\text{av}^2:=(N-1)r+s,
\end{equation}
where $\sum_{i\in V}\sum_{j\in N_C(i)}\frac{1}{N}p_{ij}=1$. Note that $|N_I(i)\cap N_I(j)|\leq N-1$ and $m_i\leq N$ which implies $T_\text{av}^1\leq T_\text{av}^2$.

In the following, we discuss the number of iterations required for each algorithm such that the players' actions converge to a Nash equilibrium.
To simplify the analysis, we assume constant step sizes (i.e., $\alpha_{k,i}=\alpha_i$). Note that for constant step sizes there exists a steady-state offset between $x(k)$ and $x^*$, see \cite{Salehisadaghiani2014Nash}. Let this minimum value of error be denoted by $d^*$, i.e., $\inf_{k}\|x(k)-x^*\|=d^*$. We use a modified \emph{$\epsilon$-averaging time} similar to Definition~1 in \cite{boyd2006randomized} for the convergence time.
\begin{definition}\label{epsilon_averaging}
	For any $0<\epsilon<1$, the $\epsilon$-averaging time of an algorithm, $N_\text{av}(\epsilon)$,  is defined as
	\begin{eqnarray}
	N_\text{av}(\epsilon)\!:=\!\sup_{x(0)}\!\inf\!\Big\{k\!:\!\text{Pr}\Big(\frac{\|x(k)\!-\!x^*\|\!-\!d^*}{\|x(0)\|}\!\geq\!\epsilon\Big)\!\leq\!\epsilon\Big\}.
	\end{eqnarray}
	\vspace{-0.5cm}
\end{definition}
By Definition~\ref{epsilon_averaging}, $N_\text{av}(\epsilon)$ is the minimum number of iterations it takes for $\|x(k)-x^*\|$ to approach an $\epsilon$-ball around $d^*$ with a high probability, regardless of $x(0)$. The following assumption guarantees $N_\text{av}(\epsilon)$ to be well-defined.
\begin{assumption}\label{non_zero_minimum_value}
	We assume a non-zero minimum value, denoted by $x_\text{min}(0)$, for the norm of the initial action of player $i$ for $i\in V$, i.e., $\|x_i(0)\|\geq x_\text{min}(0)>0$.
\end{assumption}

We obtain a lower bound for the $\epsilon$-averaging time under Algorithms~1, 2 by applying \emph{Markov's inequality}: for any random variable $X\geq0$ and $\epsilon>0$, the following holds:
\begin{equation}\label{markov_inequality}
\text{Pr}(X\geq\epsilon)\leq\frac{\mathbb{E}[X]}{\epsilon}.
\end{equation}
For constant step sizes we consider the following assumption rather than Assumption~\ref{Lip_assump}.
\begin{assumption}\label{strongly_monotone}
	$F:\Omega\rightarrow\mathbb{R}^N$ is strongly monotone on $\Omega$ with a constant $\mu>0$, i.e.,
	\begin{equation}
	(F(x)-F(y))^T(x-y)\geq\mu\|x-y\|^2\quad\forall x,y\in\Omega.
	\end{equation}
\end{assumption}
\begin{theorem}\label{theorem_convergence_rate}
	Let $\alpha_i$ be constant step sizes which satisfy $0<\phi<1$ where,
	\begin{equation}\label{condition}
	\phi:=1+(1+\rho^2+2\alpha_\text{max})p_\text{max}\alpha_{\text{max}}-(1+\rho^2+2\mu)p_{\text{min}}\alpha_{\text{min}},
	\end{equation}
	with $p_{\text{max}}=\max_{i\in V}{p_i}$, $p_{\text{min}}=\min_{i\in V}{p_i}$, $\alpha_{\text{max}}=\max_{i\in V}{\alpha_{i}}$, $\alpha_{\text{min}}=\min_{i\in V}{\alpha_{i}}$, $\rho$ be the Lipschitz constant of $F$ and $\mu$ be the positive constant for the strong monotonicity property of $F$. Under Assumptions~\ref{connected_undirected}-\ref{assump}, \ref{Lip_assump2}, \ref{non_zero_minimum_value}, \ref{strongly_monotone}, the $\epsilon$-averaging time $N_{\text{av}}(\epsilon)$ is bounded as follows:
	\begin{equation*}\label{N_av}
	N_\text{av}(\epsilon)\geq\frac{\log \frac{a}{\epsilon^{3}-b}}{\log\frac{1}{\sqrt{\gamma}}},
	\end{equation*}
	where $\gamma=\mathbb{E}\Big[\|Q(k)\|^2\Big |\mathcal{M}_k\Big]$ (as in Lemma~\ref{gamma_less_1}), $Q(k)=[(W(k)-\frac{1}{N}\textbf{1}_N\textbf{1}_N^TW(k))\otimes I_N]$, and $a, b$ are positive and increasing with $\gamma$.
\end{theorem}
\par{\emph{Proof}}. The proof follows by upper bounding $\mathbb{E}\Big[\|x(k+1)-x^*\|^2\Big]$ and then using Markov's inequality \eqref{markov_inequality} to obtain a lower bound for $N_\text{av}(\epsilon)$.
First, we start to find an upper bound for $\mathbb{E}\Big[\|\tilde{x}(k+1)-Z(k+1)\|\Big]$ which is required to upper bound $\mathbb{E}[\|x(k+1)-x^*\|^2]$. After some manipulations one can obtain,
\begin{equation}\label{error_term_squared222}
\mathbb{E}\Big[\|\tilde{x}(k+1)-Z(k+1)\|^2\Big]\leq C_1\sqrt{\gamma}^{k+1}+C_{2},
\end{equation}
where $C_1,C_2>0$ depend on $\gamma$, $\alpha_\text{max}$ and $N$. Using \eqref{error_term_squared222}, after some manipulations one can obtain,
\begin{equation}\label{error_bade_Lipschitz_expected6}
\mathbb{E}\Big[\|x(k+1)-x^*\|^2\Big]\leq C_3+C_4\sqrt{\gamma}^k,
\end{equation}
where $C_3:=\frac{\max\{Nx_{\max}^2,4NC^2p_{\max}\alpha_{\text{max}}^2+2L^2p_{\max}C_{2}\}}{1-\phi}$ and $C_4:=\frac{2L^2p_{\max}C_1}{1-\phi}$, $\phi$ is as defined in \eqref{condition}, $L=\max_{i\in V}L_i$ and $L_i$ is a Lipschitz constant (Assumption~\ref{Lip_assump2}).
Recall that $\inf_{k}\|x(k)-x^*\|=d^*$. Then,
\begin{eqnarray}\label{d*<C7}
{d^*}^2=\inf_{k}\|x(k)-x^*\|^2\leq\lim_{k\rightarrow\infty}\mathbb{E}[\|x(k)-x^*\|^2]\leq C_3.
\end{eqnarray}
Since $0<\phi<1$, by using Markov's inequality \eqref{markov_inequality} and \eqref{error_bade_Lipschitz_expected6}  the following inequality follows,
\begin{eqnarray}\label{probability_inequality}
\text{Pr}\Big(\frac{\|x(k)-x^*\|-d^*}{\|x(0)\|}\geq\epsilon\Big)
\leq \epsilon^{-2}\frac{C_4\sqrt{\gamma}^k+C_3-{d^*}^2}{\|x(0)\|^2}.
\end{eqnarray}
Using  Definition~\ref{epsilon_averaging} and Assumption~\ref{non_zero_minimum_value}, one can obtain a lower bound for $N_\text{av}(\epsilon)$ from \eqref{probability_inequality},
\begin{equation}\label{T_av_asli}
N_\text{av}(\epsilon)\geq\frac{\log \frac{a}{\epsilon^{3}-b}}{\log\frac{1}{\sqrt{\gamma}}},
\end{equation}
where $a:=C_4x_{\min}^{-2}$ and $b:=(C_3-{d^*}^2)x_{\min}^{-2}$. By Lemma~\ref{gamma_less_1}, \eqref{d*<C7} and the condition on $\phi$, $0<\phi<1$, $a$ and $b$ are positive and increasing functions of $\gamma$.
$\hfill\blacksquare$

\begin{lemma}\label{gamma=lambda2}
Let $\bar{W}:=\mathbb{E}[W(k)]$ be the expected communication matrix. Then $\bar{W}$ is doubly stochastic with $\lambda_{\max}(\bar{W})=1$. Let $\lambda_2(\bar{W})$ be second largest eigenvalue of $\bar{W}$, i.e., $\lambda_2(\bar{W}):=\max_{\lambda\neq 1}\lambda(\bar{W})$. Then $\gamma$ as defined in Lemma~\ref{gamma_less_1}, $\gamma=\lambda_{\max}\big(\mathbb{E}[Q(k)^TQ(k)]\big)$  satisfies $\gamma=\lambda_2(\bar{W})$.
\end{lemma}

By Lemma~\ref{gamma=lambda2}, $\gamma$ is equal to the second largest eigenvalue of $\bar{W}$ which can be derived as follows: $\bar{W}=I_m-\frac{\sum_{i\in V}\sum_{j\in N_C(i)}\sum_{l\in\text{ind}(i,j)}(E_l^i-E_l^j)(E_l^i-E_l^j)^T}{2\sum_{i\in V}\text{deg}_{G_C}(i)}$.
This leads us to conclude that the number of iterations is dependent on the structure of the expected communication matrix, hence the parameters associated with the interference and the communication graphs.
Note that \eqref{probability_inequality} reveals that $N_\text{av}(\epsilon)$ is not only dependent on $\gamma$ but also dependent on $\phi$ \eqref{condition}, which is a parameter associated with the cost functions.

To sum up this section, from the perspective of parameters associated with $G_I$, we can conclude that each iteration length is reduced when we consider the interference graph. Moreover, the number of iterations for each algorithm is tightly dependent on the second largest eigenvalue of the expected communication matrix hence on $G_C$.
\section{Simulation Results}
In this section we present a numerical example and compare Algorithm~1 and Algorithm~2. Consider a \emph{Wireless Ad-Hoc Network} (WANET) which consists of 16 mobile nodes interconnected by multi-hop communication paths \cite{jayash7}. Consider $N_{ah}=\{1,\ldots,16\}$ as the set of wireless nodes and $L_{ah}=\{L_l\}_{l\in\mathcal{L}}$ as the set of links connecting the nodes with $\mathcal{L}=\{1,\ldots,16\}$ as the set of link indices. Let $V=\{U_1,\ldots,U_{15}\}$ denote the set of users (players) who want to use this wireless network to transfer data. Fig.~1~(a) represents the topology of the WANET in which a unique path is assigned to each user to transfer his data from the source to the destination node. Each $U_i$ is characterized by a set of links (path), $R_i$, $i\in V$.
\begin{figure}
\vspace{-1cm}
\hspace{-2.7cm}
\centering
\includegraphics [scale=0.5]{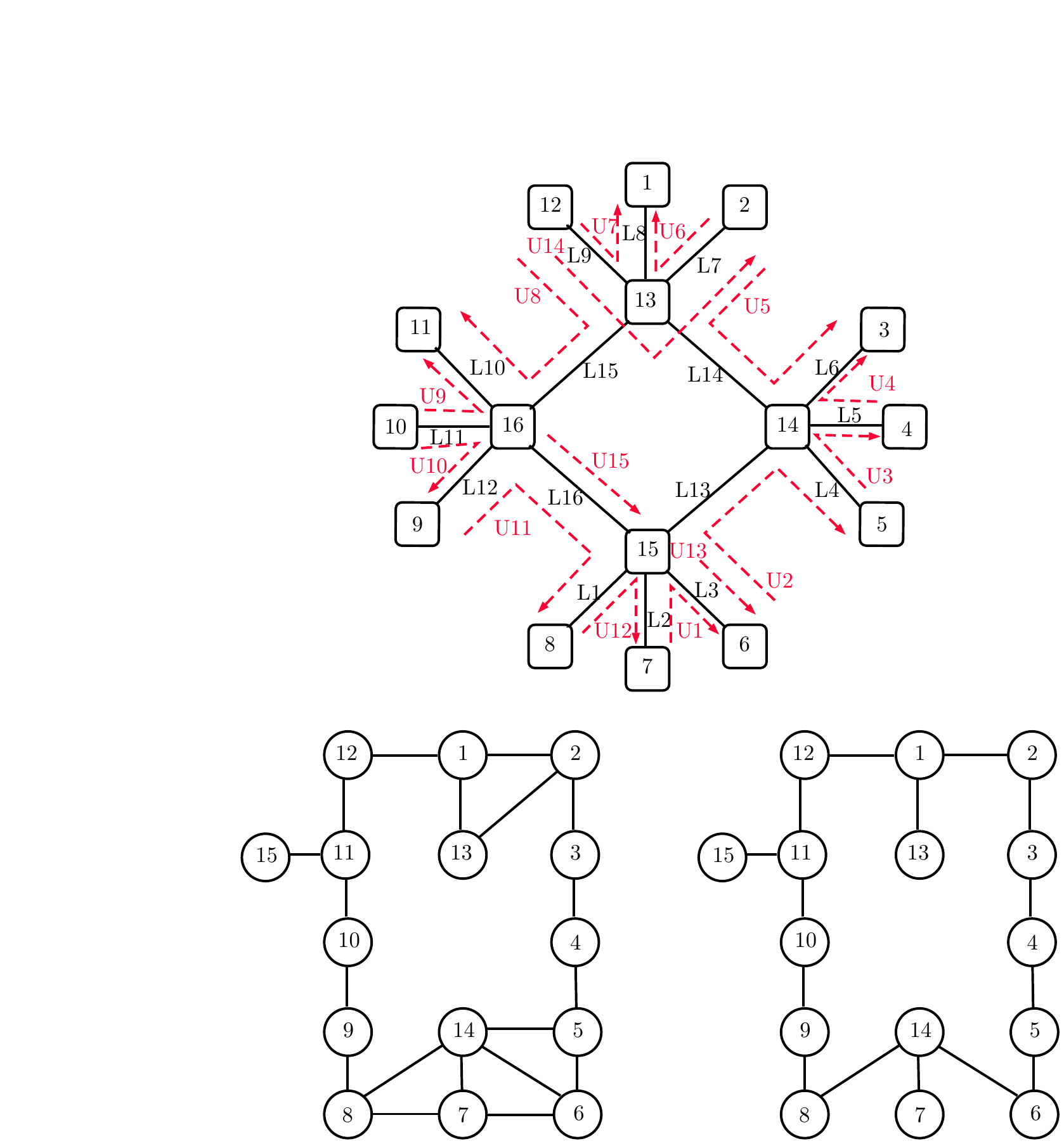}
\caption{(a)  Wireless Ad-Hoc Network. (b)  Interference graph of the Wireless Ad-Hoc Network $G_I$ (left). (c) Communication graph for the Wireless Ad-Hoc Network $G_C$ (right).}
\end{figure}

The interferences of the users to each other are represented in Fig.~1~(b). Nodes specify the users and edges demonstrate which users have a common link in their paths.
Each link $L_j\in L_{ah}$ has a positive capacity $C_j>0$ for $j\in\mathcal{L}$. Each $U_i$, $U_i\in V$, sends a non-negative flow $x_i$, $0\leq x_i\leq 10$, over $R_i$. For each $U_i$, a cost function $J_i$ is defined as
\begin{equation*}\label{Cost_fcn_gen}
J_i(x_i,x_{-i}^i):=\sum_{j:L_j\in R_i}\frac{\kappa}{C_j-\sum_{w:L_j\in R_w}x_w}-\chi_i \log(x_i+1),\end{equation*}
where $\kappa$ is a positive network-wide known parameter and $\chi_i$ is a positive user-specific parameter. The notation $a:b\in c$ translates into "set of $a$'s such that $b$ is contained in $c$".

We investigate the effectiveness of Algorithm~1 over the communication graph $G_C$ which is depicted in Fig.~1~(c). Then we compare its convergence rate with Algorithm~2 over the same $G_C$.
Let $\chi_i=10$ for $i\in V$ and $C_j=10$ for $j\in\mathcal{L}$.
\begin{figure}
\vspace{-2.5cm}
\hspace{-2.7cm}
\centering
\begin{minipage}[b]{0.45\linewidth}
\includegraphics [scale=0.31]{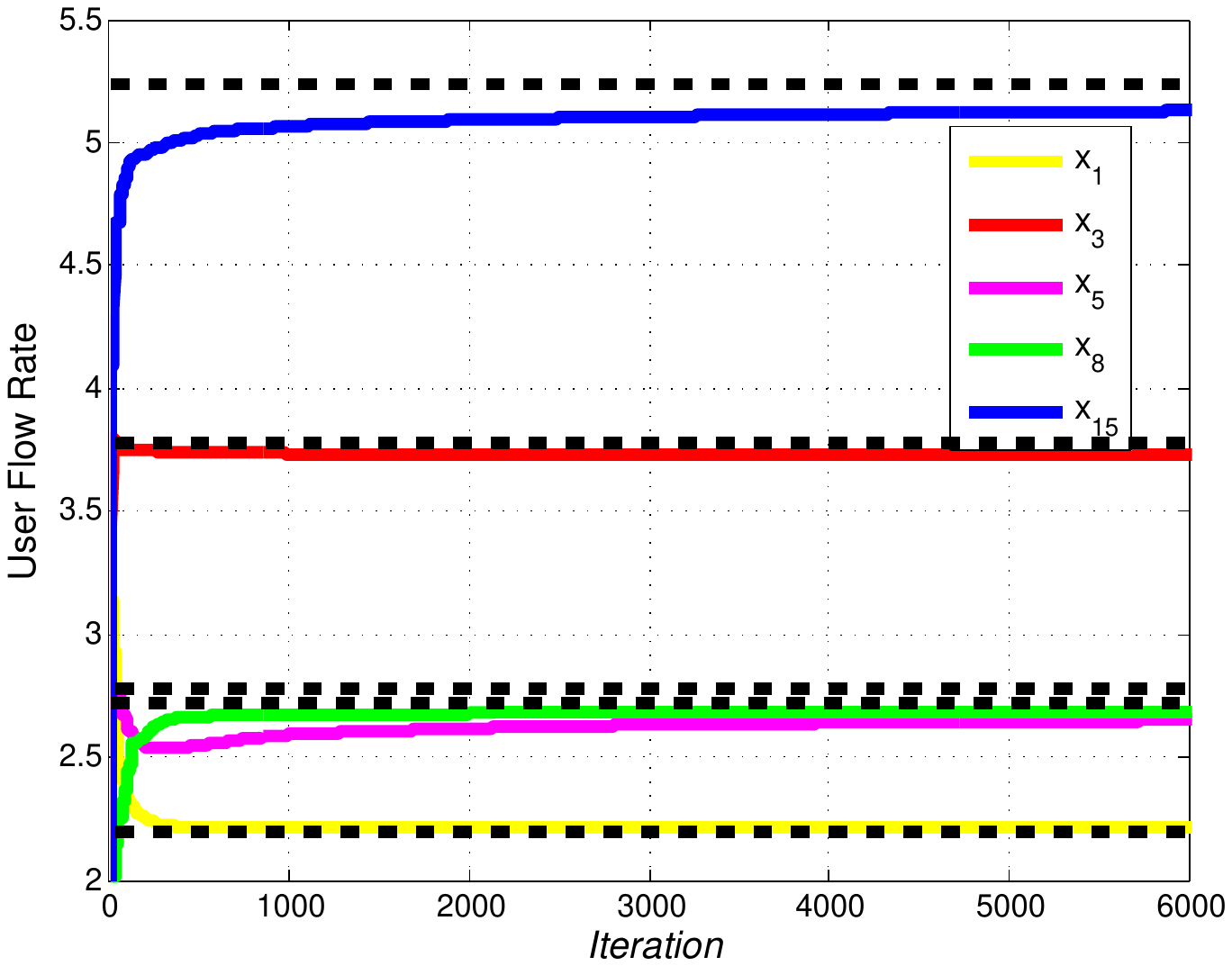}
\label{fig:minipage1}
\end{minipage}
\hspace{0.15cm}
\begin{minipage}[b]{0.45\linewidth}
\includegraphics [scale=0.31]{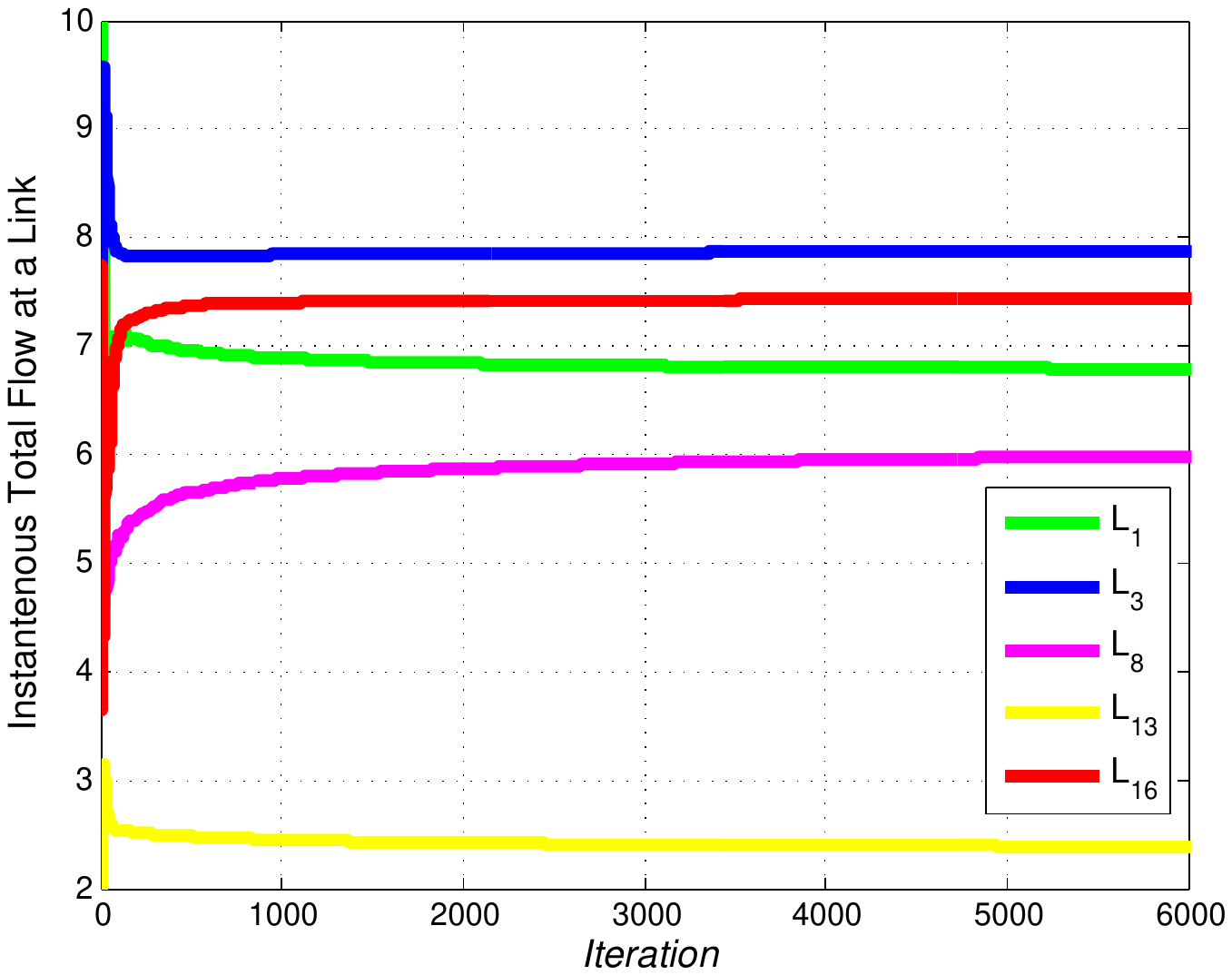}
\label{fig:minipage2}
\end{minipage}
\vspace{-3cm}
\caption{Flow rates and total flow rates (Algorithm~1).}
\vspace{-0.5cm}
\end{figure}
\begin{figure}
\vspace{-3cm}
\hspace{-2.7cm}
\centering
\begin{minipage}[b]{0.45\linewidth}
\includegraphics [scale=0.31]{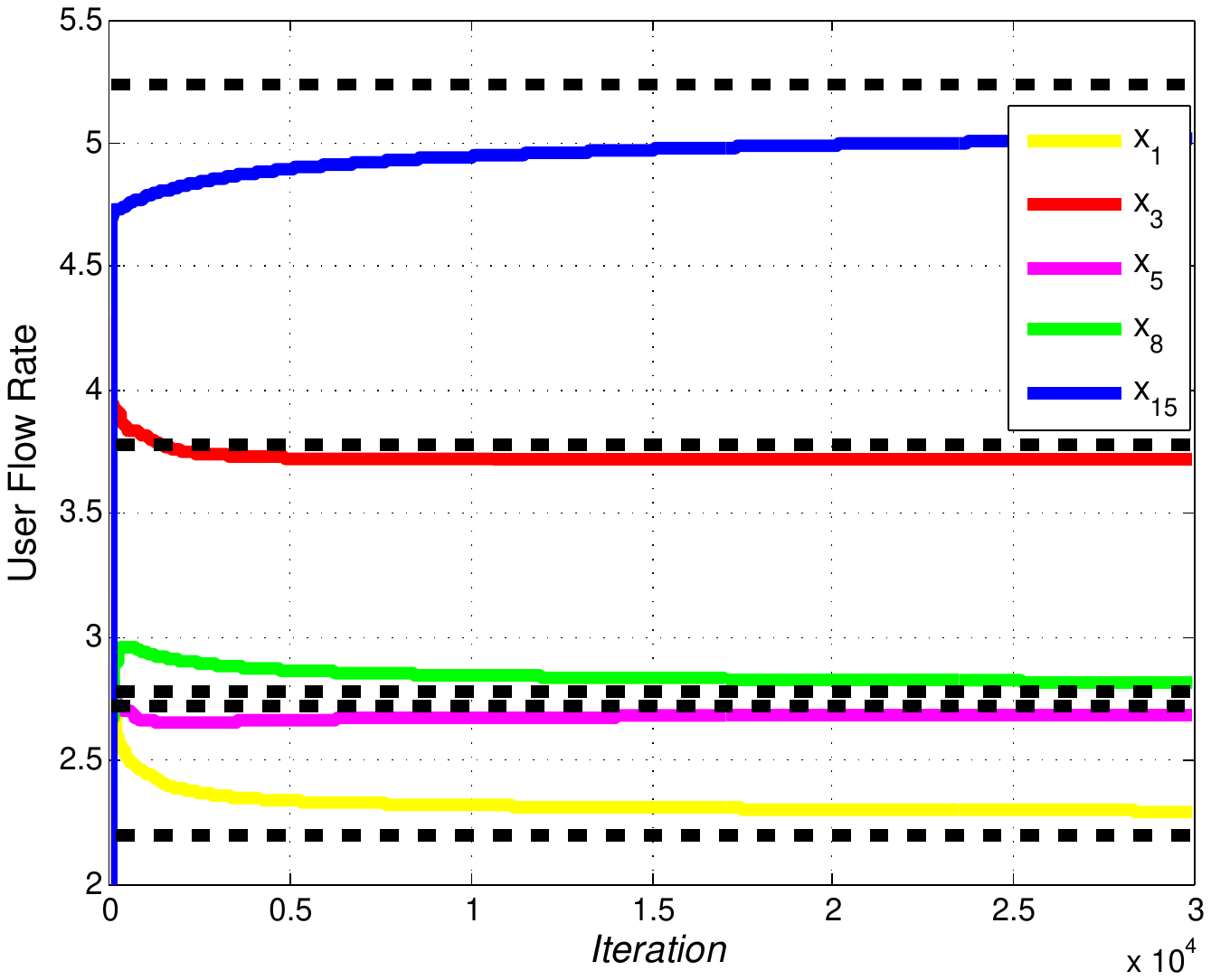}
\label{fig:minipage1}
\end{minipage}
\hspace{0.15cm}
\begin{minipage}[b]{0.45\linewidth}
\includegraphics [scale=0.31]{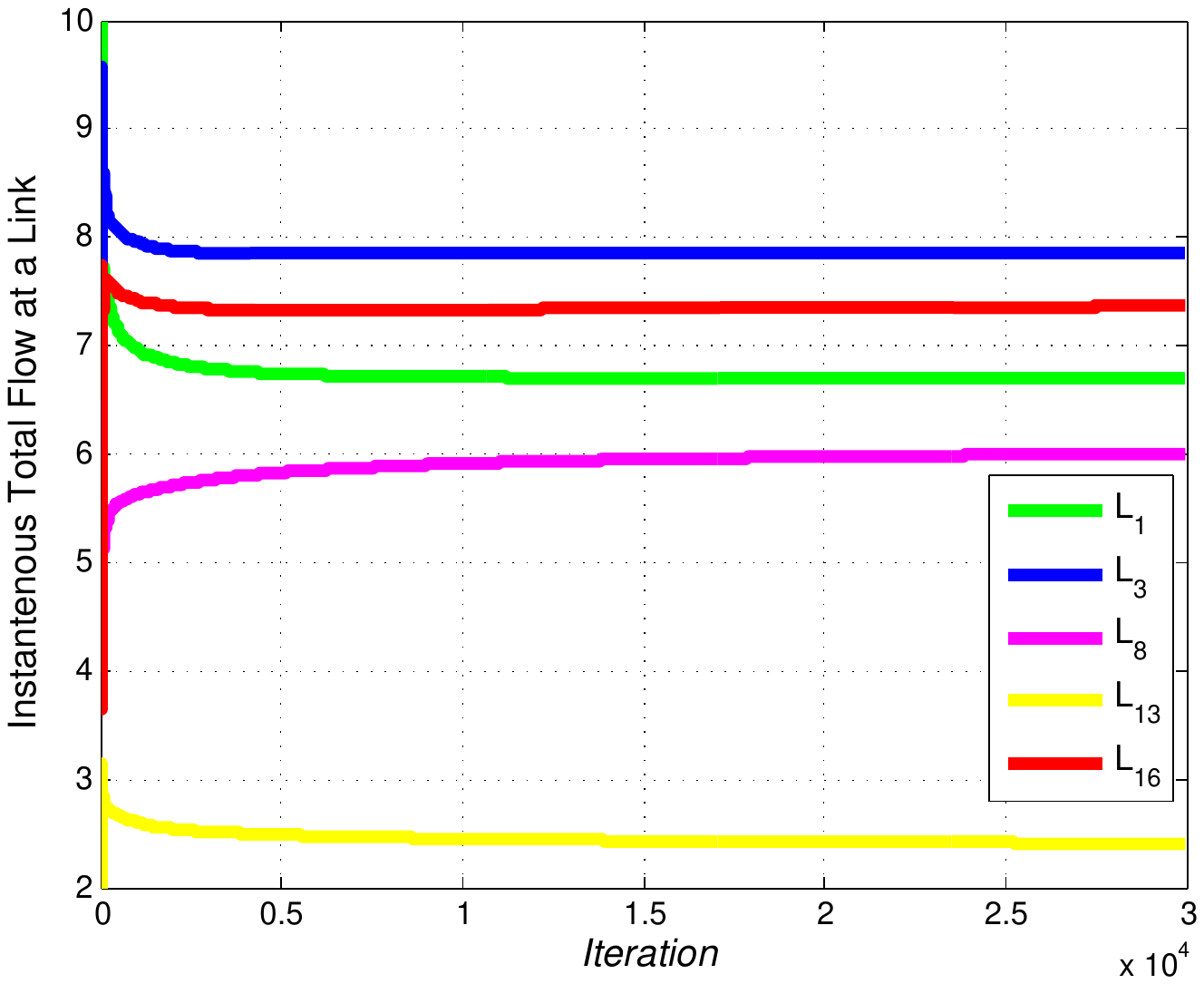}
\label{fig:minipage2}
\end{minipage}
\vspace{-3cm}
\caption{Flow rates and total flow rates (Algorithm~2).}
\end{figure}
Fig.~2 and Fig~3 show convergence of Algorithm~1 and Algorithm~2 for diminishing step sizes, respectively. The dashed lines represent the Nash equilibrium of this game. For Algorithm~1 after 6000 iterations and Algorithm~2 after 30000 iterations, the normalized error ($\frac{\|x-x^*\|}{\|x^*\|}\times100\%$) is 3.93\%. Algorithm~1 needs 5 times fewer iterations, and each iteration is 6 times shorter. Thus, Algorithm~1 is 30 times faster than Algorithm~2 in this example.

\section{Conclusions}
A gossip algorithm is proposed to find a Nash equilibrium over a network. A connected interference graph is used to illustrate the locality of the cost functions. Then, a generalized communication graph is designed for the players to exchange only their required information. Using standard assumptions on the cost functions, interference and communication
graph we proved the convergence to a Nash equilibrium. The convergence rate of the algorithm is then studied and the effect of the second largest eigenvalue of the
expected communication matrix is investigated.
\bibliographystyle{IEEEtran}
\bibliography{IEEEabrv,ref}

\begin{thebibliography}{10}
\providecommand{\url}[1]{#1}
\csname url@samestyle\endcsname
\providecommand{\newblock}{\relax}
\providecommand{\bibinfo}[2]{#2}
\providecommand{\BIBentrySTDinterwordspacing}{\spaceskip=0pt\relax}
\providecommand{\BIBentryALTinterwordstretchfactor}{4}
\providecommand{\BIBentryALTinterwordspacing}{\spaceskip=\fontdimen2\font plus
\BIBentryALTinterwordstretchfactor\fontdimen3\font minus
  \fontdimen4\font\relax}
\providecommand{\BIBforeignlanguage}[2]{{%
\expandafter\ifx\csname l@#1\endcsname\relax
\typeout{** WARNING: IEEEtran.bst: No hyphenation pattern has been}%
\typeout{** loaded for the language `#1'. Using the pattern for}%
\typeout{** the default language instead.}%
\else
\language=\csname l@#1\endcsname
\fi
#2}}
\providecommand{\BIBdecl}{\relax}
\BIBdecl

\bibitem{stankovic2012distributed}
M.~S. Stankovic, K.~H. Johansson, and D.~M. Stipanovic, ``Distributed seeking
  of {N}ash equilibria with applications to mobile sensor networks,''
  \emph{IEEE Transactions on Automatic Control}, vol.~57, no.~4, pp. 904--919,
  2012.

\bibitem{yin2011nash}
H.~Yin, U.~V. Shanbhag, and P.~G. Mehta, ``Nash equilibrium problems with
  scaled congestion costs and shared constraints,'' \emph{IEEE Transactions on
  Automatic Control}, vol.~56, no.~7, pp. 1702--1708, 2011.

\bibitem{frihauf2012nash}
P.~Frihauf, M.~Krstic, and T.~Basar, ``Nash equilibrium seeking in
  noncooperative games,'' \emph{IEEE Transactions on Automatic Control},
  vol.~57, no.~5, pp. 1192--1207, 2012.

\bibitem{jayash10}
L.~Pavel, ``An extension of duality to a game-theoretic framework,''
  \emph{Automatica}, vol.~43, no.~2, pp. 226--237, 2007.

\bibitem{li2010competitive}
H.~Li and Z.~Han, ``Competitive spectrum access in cognitive radio networks:
  Graphical game and learning,'' in \emph{Wireless Communications and
  Networking Conference (WCNC), 2010 IEEE}.\hskip 1em plus 0.5em minus
  0.4em\relax IEEE, 2010, pp. 1--6.

\bibitem{chen2012spatial}
X.~Chen and J.~Huang, ``Spatial spectrum access game: Nash equilibria and
  distributed learning,'' in \emph{Proceedings of the thirteenth ACM
  international symposium on Mobile Ad Hoc Networking and Computing}.\hskip 1em
  plus 0.5em minus 0.4em\relax ACM, 2012, pp. 205--214.

\bibitem{alpcan2004hybrid}
T.~Alpcan and T.~Ba{\c{s}}ar, ``A hybrid systems model for power control in
  multicell wireless data networks,'' \emph{Performance Evaluation}, vol.~57,
  no.~4, pp. 477--495, 2004.

\bibitem{Marden2013}
N.~Li and J.~R. Marden, ``Designing games for distributed optimization,''
  \emph{IEEE Journal of Selected Topics in Signal Processing}, vol.~7, no.~2,
  pp. 230--242, 2013.

\bibitem{Pavelbook2012}
L.~Pavel, \emph{Game theory for control of optical networks}.\hskip 1em plus
  0.5em minus 0.4em\relax Birkh\"{a}user-Springer Science, 2012.

\bibitem{nisan2007algorithmic}
N.~Nisan, T.~Roughgarden, E.~Tardos, and V.~V. Vazirani, \emph{Algorithmic game
  theory}.\hskip 1em plus 0.5em minus 0.4em\relax Cambridge University Press
  Cambridge, 2007, vol.~1.

\bibitem{kearns2001graphical}
M.~Kearns, M.~L. Littman, and S.~Singh, ``Graphical models for game theory,''
  in \emph{Proceedings of the Seventeenth conference on Uncertainty in
  artificial intelligence}.\hskip 1em plus 0.5em minus 0.4em\relax Morgan
  Kaufmann Publishers Inc., 2001, pp. 253--260.

\bibitem{tekin2012atomic}
C.~Tekin, M.~Liu, R.~Southwell, J.~Huang, and S.~H.~A. Ahmad, ``Atomic
  congestion games on graphs and their applications in networking,''
  \emph{Networking, IEEE/ACM Transactions on}, vol.~20, no.~5, pp. 1541--1552,
  2012.

\bibitem{abouheaf2014multi}
M.~I. Abouheaf, F.~L. Lewis, K.~G. Vamvoudakis, S.~Haesaert, and R.~Babuska,
  ``Multi-agent discrete-time graphical games and reinforcement learning
  solutions,'' \emph{Automatica}, vol.~50, no.~12, pp. 3038--3053, 2014.

\bibitem{bramoulle2014strategic}
Y.~Bramoull{\'e}, R.~Kranton, and M.~D'amours, ``Strategic interaction and
  networks,'' \emph{The American Economic Review}, vol. 104, no.~3, pp.
  898--930, 2014.

\bibitem{candogan2012optimal}
O.~Candogan, K.~Bimpikis, and A.~Ozdaglar, ``Optimal pricing in networks with
  externalities,'' \emph{Operations Research}, vol.~60, no.~4, pp. 883--905,
  2012.

\bibitem{zhu2016distributed}
M.~Zhu and E.~Frazzoli, ``Distributed robust adaptive equilibrium computation
  for generalized convex games,'' \emph{Automatica}, vol.~63, pp. 82--91, 2016.

\bibitem{Jayash}
J.~Koshal, A.~Nedic, and U.~V. Shanbhag, ``A gossip algorithm for aggregative
  games on graphs,'' in \emph{IEEE 51st Conference on Decision and Control
  (CDC)}, 2012, pp. 4840--4845.

\bibitem{Salehisadaghiani2014Nash}
F.~Salehisadaghiani and L.~Pavel, ``Nash equilibrium seeking by a gossip-based
  algorithm,'' in \emph{IEEE 53rd Conference on Decision and Control (CDC)},
  2014, pp. 1155--1160.

\bibitem{godsil2013algebraic}
C.~Godsil and G.~F. Royle, \emph{Algebraic graph theory}.\hskip 1em plus 0.5em
  minus 0.4em\relax Springer Science \& Business Media, 2013, vol. 207.

\bibitem{goddard1993note}
W.~Goddard and D.~J. Kleitman, ``A note on maximal triangle-free graphs,''
  \emph{Journal of graph theory}, vol.~17, no.~5, pp. 629--631, 1993.

\bibitem{Facchi24}
F.~Facchinei and J.-S. Pang, \emph{Finite-dimensional variational inequalities
  and complementarity problems}.\hskip 1em plus 0.5em minus 0.4em\relax
  Springer, 2003.

\bibitem{boyd2006randomized}
S.~Boyd, A.~Ghosh, B.~Prabhakar, and D.~Shah, ``Randomized gossip algorithms,''
  \emph{IEEE Transactions on Information Theory}, vol.~52, no.~6, pp.
  2508--2530, 2006.

\bibitem{jayash7}
T.~Alpcan and T.~Basar, ``A game-theoretic framework for congestion control in
  general topology networks,'' in \emph{IEEE 41st Conference on Decision and
  Control (CDC)}, vol.~2, 2002, pp. 1218--1224.

\end{thebibliography}
\end{document}